\documentclass[twocolumn,fleqn]{article}
\usepackage{mystyle,epsf}

\newcommand{\lsm}{\!\stackrel{\scriptscriptstyle <}{\scriptscriptstyle\sim}\!}
\newcommand{\gsm}{\!\stackrel{\scriptscriptstyle >}{\scriptscriptstyle\sim}\!}

\begin{document}

\twocolumn[
\vspace*{30mm}
\centerline{\LARGE Planar quasiperiodic Ising models}\vspace{3ex}
\centerline{\large Przemys{\l}aw Repetowicz, Uwe Grimm,
Michael Schreiber}\vspace{2ex}
\begin{footnotesize}
\centerline{\it Institut f\"{u}r Physik, 
               Technische Universit\"{a}t,
               D--09107 Chemnitz, Germany}
\end{footnotesize}
\centerline{\footnotesize August 6, 1999}\vspace{4ex}
\begin{small}
\hrule\vspace{2ex}
\begin{minipage}{\textwidth}
{\bf Abstract}\vspace{2ex}\\ 
\hp 
We investigate zero-field Ising models on periodic approximants of
planar quasiperiodic tilings by means of partition function zeros and
high-temperature expansions. These are obtained by employing a
determinant expression for the partition function. The partition
function zeros in the complex temperature plane yield precise
estimates of the critical temperature of the quasiperiodic
model. Concerning the critical behaviour, our results are compatible
with Onsager universality, in agreement with the Harris-Luck criterion
based on scaling arguments.\vspace{2ex}
\end{minipage}\vspace{2ex}
\hrule
\end{small}\vspace{6ex}
]

\section{Introduction}

\hp The recent, rather controversial discussions about magnetic
ordering in quasicrystals, in particular for alloys containing rare
earth elements \cite{COS,IFZCSG}, has revived the interest in simple
models of magnetic ordering in quasicrystals and their properties.
The influence of an aperiodic order on magnetic phase transitions can
be understood by heuristic scaling arguments \cite{H,L} resulting in
the Harris-Luck criterion for the relevance of aperiodicity. For
one-dimensional aperiodic Ising quantum chains, equivalent to
two-dimensional layered Ising models with one-dimensional
aperiodicity, exact real-space renormalization techniques can be used
that prove the heuristic criterion for a large class of models based
on substitution systems \cite{HGB,HG}. For the classical Ising model
defined on two-dimensional quasiperiodic cut-and-project tilings, the
criterion predicts that the critical behaviour is the same as for
periodic lattices, hence these models belong to the Onsager
universality class. This has unanimously been corroborated by results
obtained from various approaches \cite{GB}, and has been commonly
accepted.

Nevertheless, it is interesting to study these models in more detail,
and in this short note we present results for periodic approximants
for the Penrose and the Ammann-Beenker tiling. The zero-field Ising
model on periodic approximants can be treated by methods that allow us
to compute expansion coefficients and partition function zeros for the
{\em infinite}\/ periodic tiling, hence we do not have to deal with
finite-size corrections as one usually accounts if only finite patches
are considered. Instead, we can study the dependence on the size of
the unit cell. {}From the partition function zeros, we obtain, in
principle, the exact critical temperatures for the periodic
approximants. In this way, we are able to derive precise estimates for
the critical temperatures of the quasiperiodic models.  For the
high-temperature series of the free energy, we investigate the
fluctuations in the sequence of expansion coefficients observed in
\cite{RGS}, using the exact values for the critical temperatures of
the periodic approximants.

\section{The Ising model on periodic approximants}

\hp
We place an ``Ising spin'' $\sigma_{i}\! =\!\pm 1$ on each vertex $i$
of the periodic approximant. A configuration $\sigma\! =\!
(\sigma_1,\sigma_2,\ldots,\sigma_{n})$ is assigned the energy
\begin{equation}
E(\sigma) \; = \; -J \sum_{\langle i,j\rangle} \sigma_{i}\sigma_{j} 
\end{equation}
where $J\! >\! 0$ is the ferromagnetic exchange coupling, which we choose
identical for all pairs of neighbouring spins, and we sum over all
neighbouring pairs of spins $\langle i,j\rangle$ at vertices $i$ and
$j$. Two vertices are neighbours if they are connected by an edge of
the tiling.

\begin{figure}[t]
\centerline{\epsfxsize=0.86\columnwidth\epsfbox[90 220 495 620]{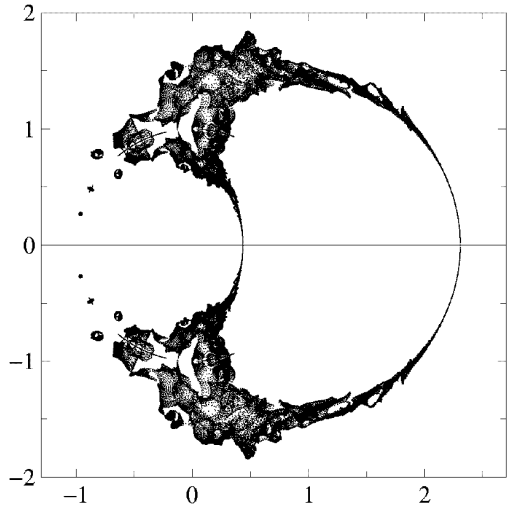}\vspace*{-1ex}}
\caption{Part of the partition function zeros in the complex 
$z$ plane for the third periodic approximant of the Penrose
tiling.\label{fig:Pen3}}
\end{figure}

There exist several variants for the solution of the zero-field Ising
model; the method we use here was initially formulated by Kac and Ward
for the square lattice Ising model \cite{KW}. The Kac-Ward determinant
and its application to periodic approximants was described in detail
in our recent paper \cite{RGS}. Let us just mention briefly that the
partition function $Z(\beta)\! =\!\sum_{\sigma}\exp[-\beta E(\sigma)]$
for a zero-field Ising model on an arbitrary planar graph, where
$\beta\! =\!1/k_{B}T$ denotes the inverse temperature, is expressed as
the square root of a determinant of an infinite matrix $K(w)$, where
$w\! =\!\tanh (\beta J)$ \cite{DZSS}. For a periodic approximant, this
matrix is cyclic. Thus the problem can, by Fourier transform, be
reduced to the computation of a determinant of a finite matrix
$\tilde{K}(w,\varphi_1,\varphi_2)$ that depends on two additional
parameters $\varphi_{1},\varphi_{2}\!\in\! [0,2\pi)$ which play the
role of Bloch phases.  The size of the matrix
$\tilde{K}(w,\varphi_1,\varphi_2)$ is given by the number of oriented
edges in the unit cell, hence by $4N$ for the Penrose and the
Ammann-Beenker approximants, where $N$ denotes the number of spins in
the unit cell. Apart from a factor $-1/\beta$, the free energy per
vertex in the thermodynamic limit $f(w)\!
=\!\lim_{M\rightarrow\infty}[\log Z(\beta)/M]$ is given as
\begin{equation}
\hspace*{-\mathindent}f(w) \; =\;
\frac{1}{2N}\!\int\limits_{0}^{2\pi}\!
\int\limits_{0}^{2\pi}\! \frac{d\varphi_1}{2\pi}\frac{d\varphi_2}{2\pi}\,
\log \det \tilde{K}(w,\varphi_1,\varphi_2)
\label{eq:f}
\end{equation}
where $M$ denotes the the number of spins in a finite patch
approximating the infinite system.

For large approximants, it is expendable, and not very profitable, to
compute the determinant analytically as a function of the parameters
$w$, $\varphi_1$, and $\varphi_2$. Instead, we rather calculate the
partition function zeros or the coefficients of the high-temperature
expansion of the free energy by exploiting the fact that the matrix
$\tilde{K}$ has the form
\begin{equation}
\tilde{K}(w,\varphi_{1},\varphi_{2}) \; = \;
I + w \tilde{L}(\varphi_{1},\varphi_{2})
\end{equation}
where $I$ denotes the unit matrix. In particular, the matrix
$\tilde{L}$ does not depend on the variable $w$. This makes it
possible to compute the partition function zeros for given values of
$\varphi_1$ and $\varphi_2$ from the eigenvalues of the matrix
$\tilde{L}(\varphi_1,\varphi_2)$, and the complete set of partition
function zeros is just the accumulation of the zeros for all
$\varphi_{1},\varphi_{2}\!\in\! [0,2\pi)$. The coefficients of the
high-temperature expansion of the free energy 
\begin{equation}
f(w)\; =\;\sum_{n=2}^{\infty}g_{2n}^{}w^{2n}
\label{eq:hte}
\end{equation}
can be calculated from traces of powers of $\tilde{L}$ by \cite{RGS}
\begin{equation}
g_{2n} = 
-\frac{1}{4nN}\!\int\limits_{0}^{2\pi}\!
\int\limits_{0}^{2\pi}\! \frac{d\varphi_1}{2\pi}\frac{d\varphi_2}{2\pi}\,
\mbox{tr}[\tilde{L}^{2n}(\varphi_{1},\varphi_{2})]
\label{eq:g}
\end{equation}
which follows by expanding the logarithm in Eq.~(\ref{eq:f}).

\section{Partition function zeros}

\begin{figure}[t]
\centerline{\epsfxsize=0.86\columnwidth\epsfbox[90 220 495 620]{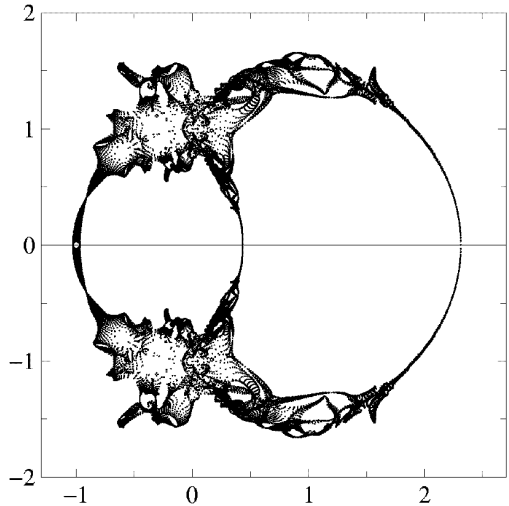}\vspace*{-1ex}}
\caption{Same as Fig.~\protect\ref{fig:Pen3}, but for the second
periodic approximant of the Ammann-Beenker tiling.\label{fig:AB2}}
\end{figure}

\hp 
In what follows, we adopt the usual convention and represent the
partition function zeros in the variable $z\! =\! (1\! +\! w)/(1\! -\!
w)\! =\!\exp(2\beta J)$.  For an infinite system, the zeros of the
partition function in the complex $z$ plane accumulate on lines or
areas that separate different regions of analyticity of the free
energy, hence different phases of the model. In particular, partition
function zeros on the real axis correspond to phase transition points,
and thus determine the critical temperature. The behaviour of zeros
around these points is governed by the corresponding critical
exponents \cite{SB}.

{}From the present approach, it is obvious that the zeros themselves
are parametrized by the two angles $\varphi_{1}$ and $\varphi_{2}$,
thus we expect that they generically fill areas in the complex plane.
In the square lattice case, however, the zeros fall on two circles
with radius $\sqrt{2}$, centred at $z\! =\!\pm 1$; similar results are
found for other simple planar lattices. In order to illustrate the
behaviour for the periodic approximants, we show the zero patterns for
the third approximant of the Penrose tiling with $N\! =\! 76$ spins
per unit cell in Fig.~\ref{fig:Pen3}, and the result for the second
approximant of the Ammann-Beenker tiling with $N\! =\! 41$ in
Fig.~\ref{fig:AB2}. Here, the zeros were computed for angles
$\varphi_{1},\varphi_{2}\!\in\!\{m\pi/20\mid m=0,1,2,\ldots,39\}$.

\begin{table}[t]
\begin{center}
\caption{Critical temperatures $w_{c}\! =\!\tanh{(J/k_{B}T_{c})}$ for 
periodic approximants of the Penrose tiling, extrapolated
to the quasiperiodic case with an estimated error. Here, $m$ labels 
the approximants with $N$ spins per unit
cell.\protect\rule[-2ex]{0pt}{2ex}\label{tab:Pen}}
\begin{tabular*}{0.67\columnwidth}{r@{\extracolsep{\fill}}r@{\extracolsep{\fill}}l}
\hline
\rule[-1.5ex]{0ex}{4.5ex}%
$m$ &
\multicolumn{1}{c}{$N$} &
\multicolumn{1}{c}{$w_{c}$} \\ \hline
\rule[0ex]{0ex}{3.25ex}%
$1$      &    $11$   & $0.401\, 440\, 380$ \\
$2$      &    $29$   & $0.395\, 411\, 099$ \\
$3$      &    $76$   & $0.395\, 082\, 894$ \\
$4$      &   $199$   & $0.394\, 554\, 945$ \\
$5$      &   $521$   & $0.394\, 523\, 576$ \\
$6$      &  $1364$   & $0.394\, 454\, 880$ \\ 
$7$      &  $3571$   & $0.394\, 451\, 035$ \\
$8$      &  $9349$   & $0.394\, 441\, 450$ \\ 
$9$      & $24476$   & $0.394\, 439\, 826$ \\
\rule[-1.5ex]{0ex}{1.5ex}%
$10$    & $64079$   & $0.394\, 439\, 319$ \\ \hline   
\rule[-1.5ex]{0ex}{4.25ex}%
$\infty$ & $\infty$ & $0.394\, 439(1)$ \\ \hline
\end{tabular*}
\vspace*{-1ex}
\end{center}
\end{table}

Clearly, the patterns are more complicated than the two circles found
for the square lattice, in particular further away from the positive
real axis. We find two zeros on the positive real axis, one for
$\varphi_{1}\! =\!\varphi_{2}\! =\! 0$ and the other for
$\varphi_{1}\! =\!\varphi_{2}\! =\!\pi$, corresponding to the
ferromagnetic and antiferromagnetic critical points, respectively,
which have the same properties due to the bipartiteness of the
tilings. In the vicinity of the critical points, the zero patterns are
very well described by segments of circles that orthogonally intersect
the real axis, showing that the corresponding critical exponent
$\alpha$ stays the same as for the square lattice, i.e.\ $\alpha\! =\!
0$, corresponding to a logarithmic singularity of the specific
heat. The zeros on the real axis determine the critical temperatures
for the periodic approximants, the results for the ferromagnetic
critical point are shown in Tables~\ref{tab:Pen} and \ref{tab:AB}. The
values are in agreement with results of recent Monte Carlo simulations
using the invaded cluster algorithm \cite{Oli}. Interestingly, the
values for the Ammann-Beenker approximants in Table~\ref{tab:AB}
appear to converge much faster than those for the Penrose case in
Table~\ref{tab:Pen}.  This might be related to the observation that
the mean coordination numbers \cite{BGJR}, in particular the number of
next-nearest neighbours, also converge faster for the Ammann-Beenker
case, thus the periodic approximants of the Ammann-Beenker tiling are,
in this sense, ``closer'' to the infinite quasiperiodic case than
those of the Penrose tiling.

\begin{table}[t]
\begin{center}
\caption{Same as Table~\protect\ref{tab:Pen}, but for periodic approximants
of the Ammann-Beenker tiling.\protect\rule[-2ex]{0pt}{2ex}\label{tab:AB}}
\begin{tabular*}{0.67\columnwidth}{c@{\extracolsep{\fill}}r@{\extracolsep{\fill}}l}
\hline
\rule[-1.5ex]{0ex}{4.5ex}%
$m$ &
\multicolumn{1}{c}{$N$} &
\multicolumn{1}{c}{$w_{c}$} \\ \hline
\rule[0ex]{0ex}{3.25ex}%
$1$      &     $7$  & $0.396\, 850\, 570$ \\
$2$      &    $41$  & $0.396\, 003\, 524$ \\
$3$      &   $239$  & $0.395\, 985\, 346$ \\
$4$      &  $1393$  & $0.395\, 984\, 811$ \\
\rule[-1.5ex]{0ex}{1.5ex}%
$5$      &  $8119$  & $0.395\, 984\, 795$ \\ \hline
\rule[-1.5ex]{0ex}{4.25ex}%
$\infty$ & $\infty$ & $0.395\, 984\, 79(2)$ \\ \hline
\end{tabular*}
\vspace*{-1ex}
\end{center}
\end{table}

\section{High-temperature expansion}

\hp In principle, the expansion coefficients of the high-temperature
series can be calculated exactly for quasiperiodic cut-and-project
tilings in the framework of the projection method.  However, the
number of graphs contributing to the expansion grows tremendously,
thus limiting the applicability of this method.  Recently \cite{RGS},
we presented the leading terms, up to 18th order in the expansion
variable $w\! =\!\tanh(\beta J)$, of the high-temperature series
(\ref{eq:hte}) for the free energy $f(w)$ of the Ising model on the
Penrose and the Ammann-Beenker tiling. The series coefficients contain
information about the critical behaviour of the model. Assuming that
the free energy $f(w)$ at the critical point $w_c$ shows a power-law
singularity $f(w)\!\sim\!{(1-w^2/w_c^2)}_{}^{\kappa}$, the ratios of
successive coefficients behave as
\begin{equation}
\frac{g_{2n}^{}}{g_{2n-2}^{}} \; \sim \; \frac{1}{w_c^2} 
\left( 1 - \frac{\kappa + 1}{n}\right)
\label{eq:rat}
\end{equation}
for large $n$, compare Eq.~(\ref{eq:hte}).  Thus, plotting the ratio
$g_{2n}/g_{2n-2}$ as a function of $1/n$, the data points should, for
sufficiently large values of $n$, lie on a straight line. {}From its
intercept at $1/n\! =\! 0$ and its slope, we can, in principle,
extract the values of the critical temperature $w_c$ and the critical
exponent $\kappa\! =\! 2-\alpha\! =\! \nu d$. Here, $d\! =\! 2$
denotes the spatial dimension, $\nu$ is the critical exponent of the
correlation length, which is $\nu\! =\! 1$ for the Onsager
universality class, and the last equality is due to scaling.

\begin{figure}[t]
\centerline{\epsfxsize=0.86\columnwidth\epsfbox{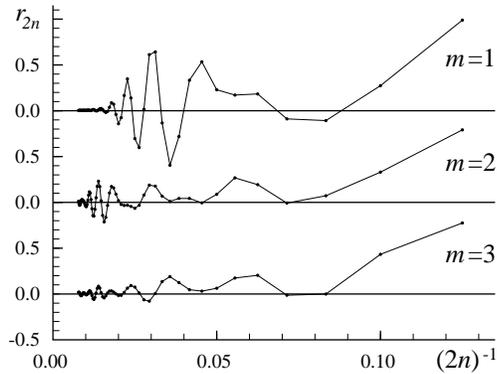}\vspace*{-1ex}}
\caption{The differences $r_{2n}$ (\protect\ref{eq:r}) between
the ratios of expansion coefficients and their expected 
asymptotic behaviour (\ref{eq:rat}) for the three smallest 
periodic approximants of the Penrose tiling. For clarity, the data 
for $m\! =\! 1$ and $m\! =\! 2$ were shifted. Lines are included to
guide the eye.\label{fig:exp1}}
\end{figure}

As shown in \cite{RGS}, this plot reveals large fluctuations, in
particular for the smallest periodic approximants, while it behaves
very nicely for simple lattices as, for instance, the square lattice.
These fluctuations rendered impossible any reasonable estimate of the
critical temperature or the critical exponent from the series
expansion, at least in this way. In order to get some more insight
into the nature and the reasons of the fluctuations, we extended the
series expansions for the periodic approximants considerably.

In Fig.~\ref{fig:exp1}, we plot the differences 
\begin{equation}
r_{2n} \; = \; 
\frac{n-3}{n\, w_c^2}  - 
\frac{g_{2n}^{}}{g_{2n-2}^{}} 
\label{eq:r}
\end{equation}
between the ratios of the expansion coefficients, up to order $2n\!
=\! 128$, and their expected asymptotic behaviour (\ref{eq:rat}), with
$\kappa\! =\! 2$, for the smallest approximants of the Penrose
tiling. Here, we use the critical temperatures $w_c$ given in
Table~\ref{tab:Pen}.  Apparently, the ratios $r_{2n}$ for the smallest
approximant with $N\! =\! 11$ spins per unit cell fluctuate strongly
for $20\lsm 2n\lsm 60$. For larger values of $2n$, the deviations from
the asymptotic behaviour are small, similar to the square lattice
case. The second approximant with $N\! =\! 29$ shows a strikingly
similar region with strong fluctuations, but at larger orders,
starting around $2n\gsm 50$ and waning just around the maximal order
$2n\! =\! 128$ in Fig.~\ref{fig:exp1}. For the third approximant, no
comparable region of particularly strong fluctuations is observed, but
we might speculate that the order where these fluctuations occur grows
roughly with the number of spins in the unit cell. In this case one
has to go to higher orders to observe the effect. One might expect
that for orders clearly beyond the number of oriented edges in the
unit cell, i.e., for $n\!\gg\! 4N$, the deviations from the asymptotic
behaviour (\ref{eq:rat}) are small.

For larger approximants, the expansion coefficients $g_{2n}$ rapidly
approach those of the quasiperiodic tilings \cite{RGS}. Thus, we may
draw the conclusion that the data for the periodic approximants show
cross-over phenomena between the characteristic behaviour for the
quasiperiodic model and the simple square-lattice case, although for
the cases considered here the asymptotic behaviours should coincide. It
seems reasonable that the cross-over occurs approximately at the size
of the unit cell.

\section{Concluding remarks}

\hp The Ising model defined on quasiperiodic tilings such as the
Ammann-Beenker and the Penrose tiling belongs, according to scaling
arguments, to the same universality class as the model on the square
lattice. This is in accordance with our results on the partition
function zeros and the high-temperature expansions for the
corresponding periodic approximants. Nevertheless, the quasiperiodic
systems show interesting features, and a better understanding
of the cross-over phenomena in the high-temperature expansions of the
periodic approximants is desirable.

\section*{Acknowledgements}

\hp 
U.~G.\ thanks M.~Baake and O.~Redner for discussions and helpful
comments.  This work has been supported by Deutsche
Forschungsgemeinschaft (DFG).

\vspace{2ex}
\begin{footnotesize}

\end{footnotesize}

\end{document}